\documentclass[12pt]{article}

\begin{document}

\title{Semiclassical Theory for Parametric
Correlation of Energy Levels}

\author{Taro Nagao$^1$, Petr Braun$^{2,3}$,
Sebastian M{\"u}ller$^4$, \\  Keiji Saito$^5$,
Stefan Heusler$^2$ and Fritz Haake$^2$}

\date{}
\maketitle

\begin{center}
\it
$^1$ Graduate School of Mathematics,
Nagoya University, Chikusa-ku, \\ Nagoya 464-8602, Japan \\
\it
$^2$ Fachbereich Physik, Universit{\"a}t Duisburg-Essen,
45117 Essen, Germany \\
\it
$^3$ Institute of Physics, Saint-Petersburg University, 198504,
Saint-Petersburg, Russia \\
\it
$^4$ Cavendish Laboratory, University of Cambridge, J J
Thomson Avenue, Cambridge CB3 0HE, UK \\
\it
$^5$ Department of Physics, Graduate School of Science,
University of Tokyo, Hongo 7-3-1, Bunkyo-ku, Tokyo 113-0033, Japan
\end{center}

\begin{abstract}

Parametric energy-level correlation describes the response of the
energy-level statistics to an external parameter such as the
magnetic field. Using semiclassical periodic-orbit theory for a
chaotic system, we evaluate the parametric energy-level
correlation depending on the magnetic field difference. The
small-time expansion of the spectral form factor  $K(\tau)$ is
shown to be in agreement with the prediction of parameter
dependent random-matrix theory to all orders in $\tau$.

\end{abstract}


\medskip


\newpage

\section{Introduction}
\setcounter{equation}{0}
\renewcommand{\theequation}{1.\arabic{equation}}

More than two decades have passed since the universal energy
level statistics was conjectured for classically chaotic
systems\cite{BGS}. Spectral correlations were found to
coincide with the predictions of Random Matrix Theory (RMT).
If the system is time-reversal invariant, the energy-level
correlation in the semiclassical limit is asymptotically
in agreement with the eigenvalue correlation
of the Gaussian Orthogonal Ensemble (GOE) of random matrices.
In a magnetic field the time-reversal invariance is broken
and the energy-level statistics is then qualitatively affected.
In that case the Gaussian Unitary Ensemble (GUE) gives a precise
prediction for the asymptotic behavior of the energy-level
correlation.

Much effort has been paid to explain the agreement with RMT in terms
of the semiclassical periodic-orbit theory\cite{MG}. A typical
physical quantity, the spectral form factor $K(\tau)$, can be written as
a sum over periodic-orbit pairs. Berry calculated the leading
contribution, of first order in the time variable $\tau$, by means of
the diagonal approximation\cite{BERRY} which is applied to both of the
GOE and GUE universality classes. For a system with time-reversal
invariance, the pairs of identical orbits and the pairs of mutually
time reversed orbits both contribute to the first-order term. For a system
without time-reversal invariance, we need to care only about the pairs
of identical orbits. In this way one is able to partially reproduce
the RMT prediction using periodic-orbit theory.

Berry's work was extended to the second-order term by Sieber and
Richter (SR) who specified the family of contributing orbit
pairs\cite{SR}. The possibility to include more complicated orbit
pairs by a combinatorial method was soon noticed. Heusler et al.
developed the analysis to the third-order term\cite{HMBH04} and M{\"u}ller
et al. obtained the expansion in agreement with the RMT result to
all orders\cite{MHBHA04,MHBHA05,SM}.

On the other hand, it is also conjectured that parameter-dependent
random matrices describe the transition of level
statistics within and in between the universality
classes\cite{LH,HAAKE}. Saito and Nagao\cite{SN} applied
semiclassical periodic-orbit theory to the parametric transition
between the GOE and GUE universality classes and obtained agreement
with ``parametric'' RMT up to the third order.  In this paper, we deal
with the parametric transition within the GUE symmetry class,
employing the magnetic field as the parameter.  Using semiclassical
periodic-orbit theory, we evaluate the small-time expansion of the
spectral form factor for the parametric correlation. The agreement
with parametric RMT is established to all orders.

This paper is organized as follows. In {\S} 2, a parametric
random-matrix theory is developed and an RMT prediction for the
spectral form factor is deduced. In {\S} 3 and {\S} 4, we employ
periodic-orbit theory for a chaotic system in a magnetic field to show
that a small-time expansion of the form factor agrees with the RMT
prediction. In {\S} 5, the key identity (a sum formula) used in {\S} 4 is
proved. In addition, a similar description of the GOE to GUE
transition is briefly given in last section.

\section{Parametric Random Matrix Theory}
\setcounter{equation}{0}
\renewcommand{\theequation}{2.\arabic{equation}}

A parameter-dependent random-matrix theory
(matrix Brownian-motion model) was first formulated by
Dyson\cite{DYSON}. He considered an ensemble of $N
\times N$ hermitian random matrices $H$ which are
close to an ``unperturbed'' hermitian matrix $H^{(0)}$.
The conditional probability distribution function
of $H$ is
given by
\begin{equation}
P(H;\sigma | H^{(0)}) \ {\rm d}H \propto
{\rm exp}\left[- \frac{{\rm Tr}\left\{ (H - {\rm e}^{-\sigma} H^{(0)})^2
\right\}}{1 - {\rm e}^{- 2 \sigma}} \right] {\rm d}H
\end{equation}
with
\begin{equation}
{\rm d}H = \prod_{j=1}^N {\rm d}H_{jj} \prod_{j<l}^N {\rm d}{\rm Re}H_{jl}
\ {\rm d}{\rm Im}H_{jl}.
\end{equation}
The parametric motion of the matrix $H$ depending on the fictitious
time parameter $\sigma$ is of interest. At the initial time $\sigma = 0$,
$H$ is equated with the hermitian matrix $H^{(0)}$. In the limit $\sigma
\to \infty$, the probability distribution function (p.d.f.)
of $H$ becomes that of the GUE
\begin{equation}
P(H;\infty | H^{(0)}) \ {\rm d}H \propto {\rm e}^{- {\rm Tr}H^2} {\rm d}H,
\end{equation}
which is independent of $H^{(0)}$.
\par
Let us denote the eigenvalues of the hermitian matrices $H$ and
$H^{(0)}$ as $x_1,x_2,\cdots,x_N$ and $x^{(0)}_1, x^{(0)}_2,\cdots,
x^{(0)}_N$, respectively. Then the p.d.f. of the eigenvalues of $H$ at
$\sigma$ (under the condition that $x_j = x^{(0)}_j$ ($j=1,2,\cdots,N$) at
$\sigma
= 0$) can be derived as
\begin{eqnarray}
& & p(x_1,x_2,\cdots,x_N;\sigma | x^{(0)}_1,x^{(0)}_2,\cdots,x^{(0)}_N)
\prod_{j=1}^N {\rm d}x_j \nonumber \\
& \propto & \prod_{j=1}^N {\rm e}^{-(x_j)^2/2 + (x^{(0)}_j)^2/2}
\prod_{j<l}^N \frac{x_j - x_l}{x^{(0)}_j - x^{(0)}_l} \det[
g(x_j,x^{(0)}_l)]_{j,l=1,2,\cdots,N} \prod_{j=1}^N {\rm d}x_j, \nonumber \\
\end{eqnarray}
where
\begin{equation}
g(x,y) = {\rm e}^{-(x^2+y^2)/2} \sum_{j=0}^{\infty}
\frac{H_j(x) H_j(y)}{\sqrt{\pi} j! 2^j} {\rm e}^{- (j + (1/2)) \sigma}
\end{equation}
with the Hermite polynomials
\begin{equation}
H_j(x) = (-1)^j {\rm e}^{x^2} \frac{{\rm d}^j}{{\rm d}x^j} {\rm e}^{-x^2}.
\end{equation}
In the limit $\sigma \to \infty$, this p.d.f. becomes the p.d.f. of the
GUE eigenvalues as
\begin{equation}
p(x_1,x_2,\cdots,x_N;\infty | x^{(0)}_1,x^{(0)}_2,\cdots,x^{(0)})
=  p_{\rm GUE}(x_1,x_2,\cdots,x_N),
\end{equation}
where
\begin{equation}
\label{pgue}
p_{\rm GUE}(x_1,x_2,\cdots,x_N)
\propto \prod_{j=1}^N {\rm e}^{- (x_j)^2} \prod_{j<l}^N |x_j - x_l|^2,
\end{equation}
as expected.
\par
Now we suppose that the initial matrix $H^{(0)}$ is a
GUE random matrix, so that the p.d.f. of
$x^{(0)}_1,x^{(0)}_2,\cdots,x^{(0)}_N$ is also
given by (\ref{pgue}). Then the transition within
the GUE symmetry class (the GUE to GUE transition)
is observed. The dynamical (density-density) correlation
function which describes the GUE to GUE transition is
defined as
\begin{equation}
\rho_d(x;\sigma|y) =  N^2 \frac{I(x;\sigma|y)}{I_0},
\end{equation}
where
\begin{eqnarray}
& & I(x_1;\sigma | x^{(0)}_1)
\nonumber \\
& = &
\int_{-\infty}^{\infty} {\rm d}x_2
\int_{-\infty}^{\infty} {\rm d}x_3 \cdots
\int_{-\infty}^{\infty} {\rm d}x_N
\int_{-\infty}^{\infty} {\rm d}x^{(0)}_2
\int_{-\infty}^{\infty} {\rm d}x^{(0)}_3 \cdots
\int_{-\infty}^{\infty} {\rm d}x^{(0)}_N
\nonumber \\
& \times &
p( x_1,x_2,\cdots,x_N;\sigma | x^{(0)}_1,x^{(0)}_2,\cdots,x^{(0)}_N)
p_{\rm GUE}(x^{(0)}_1,x^{(0)}_2,\cdots,x^{(0)}_N)
\nonumber \\
\end{eqnarray}
and
\begin{equation}
I_0 = \int_{-\infty}^{\infty} {\rm d}x
\int_{-\infty}^{\infty} {\rm d}y
I(x;\sigma|y).
\end{equation}
The dynamical correlation function describes correlations
between the spectra of $H$ and $H_0$.
\par
It is possible to evaluate the asymptotic
limit $N \to \infty$ of the dynamical
correlation function\cite{SPOHN,NF}. Introducing scaled
parameters $\eta,X,Y$ as
\begin{equation}
\sigma = \eta/(4 \pi^2 \rho^2),  \ \ \
x  = \sqrt{2 N} z + (X/\rho), \ \ \  y = \sqrt{2 N} z + (Y/\rho)
\end{equation}
($\rho = \sqrt{2 N (1 - z^2)}/\pi$ is the asymptotic
eigenvalue density at $\sqrt{2 N} z$, \ $-1< z < 1$),
we find
\begin{equation}
\label{corr}
\frac{\rho_d(x;\sigma|y)}{\rho^2} - 1 \sim {\bar \rho}(\xi;\eta) \equiv
\int_0^1 {\rm d}u \ {\rm e}^{u^2 \eta/4} \cos(\pi u \xi)
\int_1^{\infty} {\rm d}v \ {\rm e}^{- v^2 \eta/4}
\cos(\pi v \xi),
\end{equation}
where $\xi = X - Y$. The Fourier transform $K_{\rm RM}(\tau) =
\int_{-\infty}^{\infty} {\rm d}\xi \ {\rm e}^{i 2 \pi \tau \xi} \ {\bar
\rho}(\xi;\eta)$
is called the form factor. For times in the interval $0 \leq \tau \leq 1$
the form factor can be written as
\begin{eqnarray}
K_{\rm RM}(\tau)&=&  \frac{1}{2}
\int_{1 - 2 \tau}^1 {\rm d}u \ {\rm e}^{- \lambda (\tau + u)} =
\frac{{\rm e}^{-\lambda}}{\lambda} \sinh (\lambda \tau)\,,\\
\lambda &=& \eta \tau \label{lambda}\,;
\end{eqnarray}
the variable $\lambda$ was introduced here because it is the expansion
of $K_{\rm RM}(\tau)$ in powers of $\tau$ at fixed $\lambda$ which
is most naturally connected with the semiclassical periodic-orbit
theory; this expansion
\begin{equation}
\label{krm}
K_{\rm RM}(\tau) =
\tau {\rm e}^{-\lambda}
\sum_{j=0}^\infty \frac{(\lambda \tau)^{2 j}}{(2 j + 1)!}
\end{equation}
will be compared with a semiclassical result. For that purpose, we
write the expansion into the form
\begin{equation}
K_{\rm RM}(\tau) = K^{\rm diag}_{\rm RM}(\tau) + K^{\rm off}_{\rm RM}(\tau)
\end{equation}
with
\begin{equation}
K^{\rm (diag)}_{\rm RM}(\tau) = \tau {\rm e}^{-\lambda}, \ \ \
K^{\rm (off)}_{\rm RM}(\tau) = \tau {\rm e}^{-\lambda}
\sum_{j=1}^\infty \frac{(\lambda \tau)^{2 j}}{(2 j + 1)!}.
\end{equation}
In {\S} 3, we evaluate the semiclassical form factor for a chaotic system
and obtain the first-order term in agreement with $K^{\rm (diag)}_{\rm
  RM}(\tau)$.  Moreover, in {\S} 4, the semiclassical calculation is
extended to yield a result in agreement with the Laplace transform
(taken for fixed $\eta$, using (\ref{lambda}))
\begin{eqnarray}
\label{rmt} \int_0^{\infty} {\rm e}^{-q \lambda}
\left.\frac{K_{\rm RM}^{\rm
(off)}(\tau)}{\tau^2}\right|_{\tau=\lambda/\eta} {\rm d}\lambda &
= & \sum_{j=1}^{\infty} \frac{1}{(2 j + 1)!} \int_0^{\infty} {\rm
e}^{-(q + 1)\lambda} \left(\frac{\lambda}{\eta}\right)^{2 j - 1}
\lambda^{2 j}
{\rm d}\lambda \nonumber \\
& = &  \sum_{j=1}^{\infty}
\frac{1}{\eta^{2 j - 1}}
\frac{(4 j - 1)!}{(2 j + 1)!}
\frac{1}{(q + 1)^{4 j}}.
\end{eqnarray}
We thus show the agreement up to all orders.

\section{Periodic-Orbit Theory for a Chaotic System}
\setcounter{equation}{0}
\renewcommand{\theequation}{3.\arabic{equation}}

We consider a bounded quantum system with $f$ degrees of freedom
in a magnetic field $B$, assuming that the corresponding classical
dynamics is chaotic. Let us denote the energy by $E$ and each phase
space point by a $2f$ dimensional vector ${\bf x} = ({\bf q},{\bf p})$,
where $f$ dimensional vectors ${\bf q}$ and ${\bf p}$ specify the position
and momentum, respectively. In the semiclassical limit $\hbar \to
0$, the energy-level density $\rho(E;B)$ can be written in the
form
\begin{equation}
\rho(E;B) \sim \rho_{\rm av}(E) + \rho_{\rm osc}(E;B).
\end{equation}
Here $\rho_{\rm av}(E)$ is the local average of the level density
and $\rho_{\rm osc}(E;B)$ describes the fluctuation around the
average.
\par
The local average of the level density is equal to the number
of Planck cells inside the energy shell
\begin{equation}
\rho_{\rm av}(E) = \frac{\Omega(E)}{(2 \pi \hbar)^f},
\end{equation}
where $\Omega(E)$ is the volume of the energy shell.
We assume that the magnetic field is sufficiently weak 
such that the cyclotron radius is much larger than the
system size and thus the presence of the magnetic field does not
significantly change $\Omega(E)$.
\par
On the other hand, the fluctuation part is given by a sum over
the classical periodic orbits $\gamma$ as
\begin{equation}
\label{osc}
\rho_{\rm osc}(E;B) = \frac{1}{\pi \hbar} {\rm Re} \sum_{\gamma}
A_{\gamma} {\rm e}^{i (S_{\gamma}(E) + \theta_{\gamma}(B))/\hbar},
\end{equation}
where $S_{\gamma}$ is the classical action 
and $A_{\gamma}$ is the stability amplitude 
(including the Maslov phase). The phase $\theta_{\gamma}(B)$ is a function of
the magnetic field and is defined as
\begin{equation}
\theta_{\gamma}(B) = B \int_{\gamma} {\bf a}({\bf q}) \cdot {\rm d}{\bf q}
= B \int g_{\gamma}(t) \ {\rm d}t, \ \ \
g_{\gamma}(t) = {\bf a}({\bf q}_{\gamma}) \cdot
\frac{{\rm d}{\bf q}_{\gamma}}{{\rm d}t},
\end{equation}
where ${\bf a}({\bf q})$ is the gauge potential which generates
the unit magnetic field and ${\bf q}_{\gamma}(t)$ describes a
classical motion in the configuration space along the orbit
$\gamma$.
\par
In analogy with (\ref{corr}), we introduce the scaled parametric
correlation function as
\begin{eqnarray}
R(s;B,B^{\prime}) & = & \left\langle \frac{
\rho\left(E + \frac{s}{2 \rho_{\rm av}(E)};B\right)
\rho\left(E - \frac{s}{2 \rho_{\rm av}(E)};B^{\prime}\right)}{
\rho_{\rm av}(E)^2} \right\rangle - 1 \nonumber \\
& \sim & \left\langle \frac{
\rho_{\rm osc}\left(E + \frac{s}{2 \rho_{\rm av}(E)};B \right)
\rho_{\rm osc}\left(E - \frac{s}{2 \rho_{\rm av}(E)};B^{\prime}\right)}{
\rho_{\rm av}(E)^2} \right\rangle.
\end{eqnarray}
Here the angular bracket means two averages, one over the
center energy $E$ and one over a time interval much smaller
than the Heisenberg time
\begin{equation}
T_H = 2 \pi \hbar \rho_{\rm av}(E) = \frac{\Omega(E)}{(2 \pi \hbar)^{f-1}}.
\end{equation}
The form factor, namely the Fourier transform of
$R(s;B,B^{\prime})$, is then written as
\begin{eqnarray}
\label{ff}
K(\tau) & = &  \int_{-\infty}^{\infty} {\rm d}s \ {\rm e}^{i 2 \pi \tau s}
R(s;B,B^{\prime}) \nonumber \\
& \sim & \left\langle \int {\rm d}\epsilon \ {\rm e}^{i \epsilon \tau
T_H/\hbar}
\frac{\rho_{\rm osc}\left(E + \frac{\epsilon}{2};B \right)
\rho_{\rm osc}\left(E - \frac{\epsilon}{2};B^{\prime}\right)}{
\rho_{\rm av}(E)} \right\rangle.
\end{eqnarray}
Putting (\ref{osc}) into (\ref{ff}), we find
that the form factor is expressed as
a double sum over periodic orbits
\begin{equation}
K(\tau) \sim \frac{1}{T_H^2} \left\langle \sum_{\gamma,\gamma^{\prime}}
A_{\gamma} A_{\gamma^{\prime}}^* {\rm e}^{i (S_{\gamma} - S_{\gamma^{\prime}})/
\hbar} {\rm e}^{i (\theta_{\gamma}(B) - \theta_{\gamma^{\prime}}(B^{\prime}))/
\hbar} \delta\left( \tau - \frac{T_{\gamma} + T_{\gamma^{\prime}}}{2 T_H}
\right)
\right\rangle
\end{equation}
(an asterisk means a complex conjugate), where $T_{\gamma}$ and
$T_{\gamma^{\prime}}$ are the periods of the periodic orbit $\gamma$
and its partner $\gamma^{\prime}$, which ``feel'' the magnetic fields
$B$ and $B^{\prime}$, respectively. We assume that the difference between these
fields is sufficiently small so that its influence on the classical
motion can be neglected; we only have to keep the resulting difference
between the magnetic phases $\theta_\gamma(B)- 
\theta_{\gamma^{\prime}}(B^{\prime})$.
\par
Let us now denote by $\gamma_{\cal T}$ a stretch of the periodic orbit $\gamma$
whose duration ${\cal T}$ is much larger than all classical correlation
times; this stretch can coincide with the whole orbit (and then ${\cal T}$ is
the orbit period).  For times large compared to the classical scales
mentioned, successive changes of the velocity ${\rm d}{\bf
  q}_{\gamma}/{\rm d} t$ can be regarded as independent random
events\cite{RICHTER}, so that a replacement of $g_{\gamma}(t)$ by
Gaussian white noise is justified. An average of a functional
$F[g_{\gamma_{\cal T}}]$ over  Gaussian white noise is evaluated as
\begin{equation}
\label{gaussian} \langle \langle \ F[g_{\gamma_{\cal T}}] \ \rangle
\rangle = \frac{\displaystyle \int {\cal D}g_{\gamma} \ {\rm
exp}\left[ -\frac{1}{4 D} \int_0^{\cal T} {\rm d}t (g_{\gamma}(t))^2
\right] F[g_{\gamma}]} {\displaystyle \int {\cal D}g_{\gamma} \
{\rm exp}\left[ -\frac{1}{4 D} \int_0^{\cal T} {\rm d}t (g_{\gamma}(t))^2
\right]}
\end{equation}
and implies a correlation $\langle \langle g_{\gamma}(t)
g_{\gamma}(t^{\prime}) \rangle \rangle = 2D \delta 
(t-t^{\prime})$. Including this
Gaussian average (carried over the whole duration of the periodic
orbits), we rewrite the form factor as
\begin{equation}
K(\tau) \sim \frac{1}{T_H^2} \left\langle \sum_{\gamma,\gamma^{\prime}}
A_{\gamma} A_{\gamma^{\prime}}^* {\rm e}^{i (S_{\gamma} - S_{\gamma^{\prime}})/
\hbar} \langle\langle
{\rm e}^{i (\theta_{\gamma}(B) - \theta_{\gamma^{\prime}}(B^{\prime}))/
\hbar} \rangle \rangle
\delta\left( \tau - \frac{T_{\gamma} + T_{\gamma^{\prime}}}{2 T_H} \right)
\right\rangle.
\end{equation}
We shall evaluate the small-$\tau$ expansion
of this semiclassical form factor, restricting
ourselves to homogeneously hyperbolic systems
with two degrees of freedom ($f=2$).
\par
Let us begin with adapting Berry's diagonal
approximation\cite{BERRY} to correlations between two spectra
pertaining to different values of the magnetic field. In this
approximation, one first considers the contributions of
periodic-orbit pairs $\gamma^{\prime} = \gamma$. 
The key ingredient is Hannay and
Ozorio de Almeida (HOdA)'s sum rule\cite{HODA}
\begin{equation}
\frac{1}{T_H^2} \sum_{\gamma}\left| A_{\gamma}\right|^2
\delta \left( \tau - \frac{T_{\gamma}}{T_{H}} \right)
= \tau.
\end{equation}
Using this sum rule and the Gaussian average (\ref{gaussian}) for
 pairs of identical orbits $({\gamma,\gamma})$ we find
\begin{equation}
\frac{1}{T_H^2} \sum_{\gamma}\left| A_{\gamma}\right|^2 \delta
\left( \tau - \frac{T_{\gamma}}{T_{H}}\right) \langle \langle {\rm
e}^{ i( \theta_{\gamma}(B) - \theta_{\gamma}(B^{\prime}))/\hbar}
\rangle \rangle = \tau {\rm e}^{- a T}.
\end{equation}
Here $T$ is the period $\tau T_H$. Since the Heisenberg time
$T_H$ is of the order $1/\hbar$ and $a = (B -
B^{\prime})^2 D/\hbar^2$ the decay rate at $\tau$ fixed is
proportional to $(B-B^{\prime})^2/\hbar^3$. The contribution of pairs of
identical orbits does not vanish in the limit $\hbar\to 0$
provided the field difference is scaled such that this parameter
remains finite.
\par
Consider now the case when the magnetic field is so weak that its
influence on the orbital motion can be neglected. Then the system
is close to being time-reversal invariant, and its periodic orbits 
occur in almost mutually time-reversed pairs $(\gamma,{\bar \gamma})$; 
these must be taken into account as well. However we can check that the 
pair $(\gamma,{\bar \gamma})$ yields no contribution. This will be 
true if both $B$ and $B^{\prime}$ are quantum mechanically large in the sense
\begin{equation}
\label{strongfield}
B,B^{\prime}  \gg  O(\hbar^{3/2}),
\end{equation}
which does not prevent the field difference from being quantum
mechanically small. Namely, as the phase factor $\theta_{\gamma}$
changes sign under time reversal,
\begin{equation}
\frac{1}{T_H^2} \sum_{\gamma}\left| A_{\gamma}\right|^2
\delta \left( \tau - \frac{T_{\gamma}}{T_{H}} \right)
\langle \langle {\rm e}^{ i( \theta_{\gamma}(B)
- \theta_{{\bar \gamma}}(B^{\prime}))/\hbar} \rangle \rangle
=\tau \langle \langle {\rm e}^{i (\theta_{\gamma}(B)
+ \theta_{\gamma}(B^{\prime}))/\hbar } \rangle \rangle
\to 0
\end{equation}
in the limit $\hbar \to 0$. It means that pairs of  time reversed
orbits do not contribute to the form factor if (\ref{strongfield})
holds.

Putting the above results together, we obtain the diagonal
approximation of the form factor as
\begin{equation}
K_{\rm PO}^{\rm (diag)} = \tau {\rm e}^{- a T}.
\end{equation}
This is in agreement with the first-order term of the RMT
prediction ({\ref{krm}), if the RMT parameter $\lambda$ is
identified with $a  T$.

\section{Off-diagonal Contributions}
\setcounter{equation}{0}
\renewcommand{\theequation}{4.\arabic{equation}}

We are now in a position to calculate the off-diagonal
contribution. In order to identify the family of periodic-orbit
pairs responsible for the leading off-diagonal terms, we note the
fact that long periodic orbits have close self-encounters where
two or more orbit segments come close in phase space. The
duration of the relevant self-encounters are of the order of the
Ehrenfest time $T_E$\cite{MHBHA05}. Although $T_E$ is
logarithmically divergent in the limit $\hbar \to 0$, it is
still vanishingly small compared to the period (which is of the
order of the Heisenberg time $T_H$). After leaving a
self-encounter, the orbit goes along a loop in phase space and
comes to a different (or back to the same) encounter. All
off-diagonal terms arise from the existence of orbits $\gamma$
which are close but different from the partners $\gamma^{\prime}$
in the encounters but almost identical to them on the loops.
Within the encounters the orbits $\gamma$ and $\gamma^{\prime}$
are differently connected to the loops. Suppose that
the magnetic fields $B$ and $B^{\prime}$ are sufficiently
strong. Then, since we are treating a system without time-reversal
invariance, $\gamma$ and its partner $\gamma^{\prime}$ go in the same
direction on all loops.
\par
Let us consider such a periodic-orbit pair $\alpha =
(\gamma,\gamma^{\prime})$ with $L$ loops and $V$
encounters. Inside each encounter, we introduce
a Poincar{\'e} section ${\cal P}$ transversal to
the orbit $\gamma$ in  phase space. Pairwise
normalized vectors ${\hat e}_s$ and ${\hat e}_u$
span the section ${\cal P}$. Here the vectors
${\hat e}_s$ and ${\hat e}_u$ have directions
along the stable and unstable manifolds, respectively.
Each segment of the orbit within the encounter
pierces through ${\cal P}$ at one phase-space
point. The displacement $\delta {\bf x}$ between
such points can be decomposed as $\delta {\bf x}
= s {\hat e}_s + u {\hat e}_u$. If we fix one
reference piercing point as the origin, each of
the others is specified by a coordinate pair
$(s,u)$.
\par
Suppose that the periodic orbit $\gamma$
pierces ${\cal P}$ within the $r$-th
encounter. If $l_r$ segments of $\gamma$
are contained in the encounter, there
are $l_r$ piercing points so that $l_r - 1$
coordinate pairs relative to the
reference piercing are necessary to
specify them. Consequently, we need
$\sum_{r=1}^V (l_r - 1) = L - V$ coordinate
pairs $(s_j,u_j)$ to specify all the
piercing points within the encounters.
\par
We denote the time elapsed on the $j$'th loop by $T_j$ and the
duration of the $r$-th encounter by $t_{{\rm enc},r}$. It
follows that the total duration of the encounters is
\begin{equation}
t_{\alpha} \equiv \sum_{r=1}^V l_r t_{{\rm enc},r}\,.
\end{equation}
Using these notations, we can employ ergodicity to estimate
the number of encounters in a periodic orbit with a period
$T = \sum_{j=1}^L T_j + t_{\alpha}$ as\cite{MHBHA04,MHBHA05,SM,SN}
\begin{equation}
\int {\rm d}{\bf u}{\rm d}{\bf s}
\int_0^{T-t_{\alpha}} {\rm d}T_1
\int_0^{T-t_{\alpha}-T_1} {\rm d}T_2 \cdots
\int_0^{T-t_{\alpha}-T_1-T_2-\cdots-T_{L-2}} {\rm d}T_{L-1}
\ Q_{\alpha},
\end{equation}
where
\begin{equation}
Q_{\alpha} = N({\vec v}) \frac{T}{L \
\prod_{r=1}^V t_{{\rm enc},r} \ \Omega^{L-V}}
\end{equation}
and the integration measures are given by
\begin{equation}
{\rm d}{\bf u} = \prod_{j=1}^{L-V} {\rm d}u_j,
\ \ \ {\rm d}{\bf s} = \prod_{j=1}^{L-V} {\rm d}s_j.
\end{equation}
The combinatorial factor $N({\vec v})$ is the
number of structures of orbit pairs for a given
vector ${\vec v} = (v_2,v_3,v_4,\cdots)$, where
the component $v_l$ denotes the number of
the encounters with $l$ segments; we will
occasionally write
\begin{equation}
{\vec v} = (2)^{v_2} (3)^{v_3} (4)^{v_4} \cdots.
\end{equation}
It should be noted that
\begin{equation}
L = \sum_{l=2}^{\infty} l v_l, \ \ \ V = \sum_{l=2}^{\infty} v_l.
\end{equation}
For $n = L-V+1 = 3$ and $5$, we tabulate
$N({\vec v})$'s in Table 1. The precise meaning
of the orbit structure is expounded in next section.
\par
We then calculate the Gaussian average (\ref{gaussian}) on the loops
and obtain a factor ${\rm e}^{-a T_1} {\rm e}^{-a T_2} \cdots {\rm e}^{-a
  T_L}$. Similarly, an encounter contributes a factor ${\rm e}^{- a
  (l_r)^2 t_{{\rm enc},r}}$.
\par
It is now straightforward to obtain the contribution
to the form factor from the orbit pair $\alpha$
\begin{eqnarray}
& & K_{{\rm PO},\alpha}(\tau)
= \tau \int {\rm d}{\bf u}{\rm d}{\bf s}
\nonumber \\ & \times  &
\int_0^{T-t_{\alpha}} {\rm d}T_1
\int_0^{T-t_{\alpha}-T_1} {\rm d}T_2 \cdots
\int_0^{T-t_{\alpha}-T_1-T_2-\cdots-T_{L-2}} {\rm d}T_{L-1}
Q_{\alpha} R_{\alpha} {\rm e}^{i \Delta S/\hbar}, \nonumber \\
\end{eqnarray}
where
\begin{equation}
R_{\alpha} = {\rm e}^{-a(T_1 + T_2 + \cdots + T_L)}
{\rm e}^{-a ((l_1)^2 t_{{\rm enc},1}
 + (l_2)^2 t_{{\rm enc},2} + \cdots +
(l_V)^2 t_{{\rm enc},V})}.
\end{equation}
The action difference $\Delta S \equiv S_{\gamma} -
S_{\gamma^{\prime}}$ is estimated as $\Delta S =
\sum_{j=1}^{L-V} u_j s_j$\cite{MHBHA04,MHBHA05,SM}. This
formula contributes to the terms of order $\tau^n$
with $n = L-V+1$.
\par
Then we expand $K_{\rm PO,\alpha}(\tau)$ in $t_{{\rm enc},r}$ and extract
the term where all $t_{{\rm enc},r}$'s mutually cancel.  Because of
the appearances of extra factors $\hbar$ or rapid oscillations in the
limit $\hbar \to 0$, the other terms give no
contribution\cite{MHBHA04,MHBHA05,SM}. We thus obtain the off-diagonal
term of the form factor
\begin{eqnarray}
K_{\rm PO}^{\rm (off)}(\tau)
& = & \sum_{{\vec v}} N({\vec v})
\frac{\tau^2}{L} \left( \frac{1}{T_H} \right)^{L-V-1}
\prod_{r=1}^V \left( - l_r \frac{\partial}{\partial T} -
(l_r)^2 a \right) f(T) \nonumber \\
 & = & \sum_{{\vec v}} N({\vec v})
\frac{\tau^2}{L} \left( \frac{1}{T_H} \right)^{L-V-1}
\prod_{l=2}^{\infty} \left( - l \frac{\partial}{\partial T} -
l^2 a \right)^{v_l} f(T),
\end{eqnarray}
where
\begin{equation}
f(T) =  \int_0^T {\rm d}T_1
\int_0^{T-T_1} {\rm d}T_2 \cdots
\int_0^{T-T_1-T_2-\cdots-T_{L-2}} {\rm d}T_{L-1}
{\rm e}^{-a(T_1 + T_2 + \cdots T_L)}.
\end{equation}
Let us put $\lambda = a T$ and calculate the Laplace
transform of $K_{\rm PO}^{\rm (off)}(\tau)/\tau^2$ as
\begin{eqnarray}
& & \int_0^{\infty} {\rm e}^{- q \lambda}
\frac{K_{\rm PO}^{\rm (off)}(\tau)}{\tau^2}
{\rm d}\lambda  \nonumber \\
& =  & \sum_{{\vec v}} N({\vec v})
\frac{1}{L} \left( \frac{1}{T_H} \right)^{L-V-1}
\int_0^{\infty} {\rm d}\lambda {\rm e}^{- q \lambda}
\prod_{l=2}^{\infty} \left( - l \frac{\partial}{\partial T} -
l^2 a \right)^{v_l} f(T) \nonumber \\
& = & \sum_{{\vec v}} N({\vec v})
\frac{a}{L} \left( \frac{1}{T_H} \right)^{L-V-1}
\prod_{l=2}^{\infty} \left( - l a q - l^2 a
\right)^{v_l} \frac{1}{(a q + a)^L} \nonumber \\
& = &
\sum_{n=2}^{\infty} \frac{1}{(q + 1)^{n-1}}
\left( \frac{1}{a T_H} \right)^{n-2}
\sum_{{\vec v}}^{L-V+1=n} {\tilde N}({\vec v})
\prod_{l=2}^{\infty} \left(1 +
(l - 1) \frac{1}{q + 1} \right)^{v_l}, \nonumber \\
\end{eqnarray}
where ${\tilde N}({\vec v}) = N({\vec v})
(-1)^V \prod_{l=2}^{\infty} l^{v_l}/L$. In the above
equation, a simple graphical rule is observed: each loop contributes
a factor $1/(a(q + 1))$ and each encounter contributes $-la(q+l)$.
In next section, we shall prove a sum formula for $n \geq 2$
\begin{equation}
\label{sf1}
\sum_{{\vec v}}^{L-V+1=n} {\tilde N}({\vec v})
\prod_{l=2}^{\infty} \left( 1 + (l - 1) \frac{1}{q + 1} \right)^{v_l} = \left\{
\begin{array}{ll} \displaystyle
\frac{(2 n - 3)!}{n!} \left( \frac{1}{q + 1}
\right)^{n-1}, & n \ {\rm odd},
\\ 0, & n \ {\rm even}, \end{array} \right.
\end{equation}
from which it follows that
\begin{equation}
\int_0^{\infty} {\rm e}^{- q \lambda}
\frac{K_{\rm PO}^{\rm (off)}(\tau)}{\tau^2}
{\rm d}\lambda = \sum_{j=1}^{\infty}
\frac{1}{(a T_H)^{2 j -1}}
\frac{(4 j - 1)!}{(2 j + 1)!}
\frac{1}{(q + 1)^{4 j}}.
\end{equation}
As $a T_H = (a T)(T_H/T) = \lambda/\tau = \eta$,
this is in agreement with the RMT result (\ref{rmt}).

\section{A Sum Formula for ${\tilde N}({\vec v})$}
\setcounter{equation}{0}
\renewcommand{\theequation}{5.\arabic{equation}}

In this section we shall give a proof for
the sum formula (see (\ref{sf1}))
\begin{equation}
\label{sf2}
\sum_{{\vec v}}^{L-V+1=n} {\tilde N}({\vec v})
\prod_{l=2}^{\infty} ( 1 + (l - 1) x )^{v_l} = \left\{
\begin{array}{ll} \displaystyle
\frac{(2 n - 3)!}{n!} x^{n-1}, & n \ {\rm odd},
\\ 0, & n \ {\rm even} \end{array} \right.
\end{equation}
with $n \geq 2$. For that purpose we introduce a
number $N_P({\vec v})$ depending on the vector
\begin{equation}
{\vec v} = (1)^{v_1} (2)^{v_2} (3)^{v_3} (4)^{v_4} \cdots
\end{equation}
and set $L = \sum_{l=1}^{\infty} l v_l$ and $V = \sum_{l=1}^{\infty}
v_l$. Let us denote an ``encounter'' permutation of the numbers
$1,2,\cdots,L$ as
\begin{equation}
P_{\rm enc} = \left( \begin{array}{ccccc} 1 & 2 & 3 & \cdots & L \\
P_{\rm enc}(1) & P_{\rm enc}(2) & P_{\rm enc}(3)
& \cdots & P_{\rm enc}(L) \end{array} \right)
\end{equation}
and define a ``loop'' permutation
\begin{equation}
P_{\rm loop} = \left( \begin{array}{cccccc} 1 & 2 & 3 & \cdots & L-1 & L \\
2  & 3 & 4 & \cdots & L & 1 \end{array} \right).
\end{equation}
We define $N_P({\vec v})$ as the number
of permutations $P_{\rm enc}$ which satisfy the
following two conditions.
\par
\medskip
\noindent
(A) The permutation $P_{\rm enc}$ has $v_l$ cycles
of length $l$.
\par
\medskip
\noindent
(B) The product $P_{\rm loop} P_{\rm enc}$ is a permutation
with a single cycle.
\par
\medskip
\noindent
Then it follows that
\begin{equation}
N((2)^{v_2} (3)^{v_3} (4)^{v_4}
\cdots ) = N_P((1)^0 (2)^{v_2} (3)^{v_3} (4)^{v_4} \cdots).
\end{equation}
In order to explain the reason, let us suppose the following
situation. The encounters include $\sum_{r=1}^V l_r = L$ orbit
segments in total, so that there are $L$ ``entrances'' where the orbits
come in and $L$ ``exits'' where the orbits go out. A periodic orbit
$\gamma$ comes in an encounter at the first ``entrance'' and goes out
at the first ``exit''. Then it comes to the second ``entrance'' and
goes out at the second ``exit''. It continues to follow the
connection pattern
\par
\medskip
\noindent
$j$-th ``entrance'' $\to$ $j$-th ``exit'' $\to$
$(j+1)$-th ``entrance''
\par
\medskip
\noindent
and finally goes out at the $L$-th ``exit'' and then comes back to the
first ``entrance'' again. On the other hand, the partner orbit
$\gamma^{\prime}$ comes in an encounter at the first ``entrance''
and goes out at $P_{\rm enc}(1)$-th ``exit''. Then it must go to
the $P_{\rm loop} P_{\rm enc}(1)$-th ``entrance'', as the partners
go along the same loop. It continues to follow the pattern
\par
\medskip
\noindent
$j$-th ``entrance'' $\to$ $P_{\rm enc}(j)$-th ``exit'' $\to$
$P_{\rm loop} P_{\rm enc}(j)$-th ``entrance''
\par
\medskip
\noindent
In this manner, if a permutation $P_{\rm enc}$ is given,
the structure of a periodic orbit $\gamma^{\prime}$ is
specified.
\par
The $j$-th ``entrance'' and the $l$-th ``exit'' belong to the
same encounter, if and only if $j$ and $l$ are contained
in the same cycle of the permutation $P_{\rm enc}$. Hence
the condition (A) is required. The orbit $\gamma^{\prime}$
finally comes to $(P_{\rm loop} P_{\rm enc})^L(1)$-th ``entrance''.
As $\gamma^{\prime}$ is a connected periodic orbit, it must
be the first return to the first ``entrance''. This is guaranteed
by the condition (B).
\par
A combinatorial argument\cite{MHBHA04,MHBHA05,SM} yields a
recursion relation for
\begin{equation}
{\tilde N}_P({\vec v}) = N_P({\vec v})
(-1)^V \prod_{l=1}^{\infty} l^{v_l}/L
\end{equation}
as
\begin{eqnarray}
\label{recur}
& & v_l {\tilde N}_P({\vec v}) + \sum_{k \geq 1}
v^{[k,l \to k + l - 1]}_{k+l-1} k {\tilde N}_P(
{\vec v}^{[k,l \to k+l-1]}) \nonumber \\
& + & \sum_{1 \leq m \leq l-2} (v_{l-m-1} + 1) v_m^{[l \to m,l-m-1]}
{\tilde N}_P({\vec v}^{[l \to m,l-m-1]}) = 0.
\end{eqnarray}
Here we used a notation
\begin{equation}
{\vec v}^{[\alpha_1,\cdots,\alpha_{\nu} \to \beta_1,\cdots,
\beta_{\nu^{\prime}}]},
\end{equation}
which is the vector obtained from ${\vec v}$ when we decrease
each of $v_{\alpha_1},v_{\alpha_2},\cdots,v_{\alpha_{\nu}}$ by one and
increase each of $v_{\beta_1},v_{\beta_2},\cdots,v_{\beta_{\nu^{\prime}}}$ 
by one.  It should be noted that ${\tilde N}_P({\vec v})$ is zero if
any of the components of ${\vec v}$ is negative.
\par
In the special case $l = 2$, we obtain a simplified recursion
formula for ${\tilde N}
({\vec v})$
\begin{equation}
\label{simp}
v_2 {\tilde N}({\vec v}) + \sum_{k \geq 2}
v_{k+1}^{[k,2 \to k+1]} k {\tilde N}({\vec v}^{[k,2 \to k+1]}) = 0.
\end{equation}
Let us introduce a variable $x$ and define
\begin{equation}
\label{nvx}
{\tilde N}({\vec v},x)
= {\tilde N}({\vec v}) \prod_{l=2}^{\infty} (1+ (l-1)x)^{v_l}.
\end{equation}
Then the recursion formula (\ref{simp}) reads
\begin{equation}
\frac{v_2}{1 + x} {\tilde N}({\vec v},x) + \sum_{k \geq 2}
\frac{k (1 + (k - 1) x)}{1 + k x} v_{k+1}^{[k,2 \to k+1]}
{\tilde N}({\vec v}^{[k,2 \to k+1]},x) = 0.
\end{equation}
Summing this over ${\vec v}$ with fixed $L-V+1=n$, we find
\begin{equation}
\sum_{{\vec v}}^{L-V+1=n} \left[ \frac{v_2}{1 + x}
{\tilde N}({\vec v},x) + \sum_{k \geq 2}
\frac{k (1 + (k - 1) x)}{1 + k x} v_{k+1}^{[k,2 \to k+1]}
{\tilde N}({\vec v}^{[k,2 \to k+1]},x) \right] = 0.
\end{equation}
Here the sum over ${\vec v}$ can be replaced by
the sum over ${\vec v}^{\prime} \equiv
{\vec v}^{[k,2 \to k+1]}$, so that
\begin{equation}
\sum_{{\vec v}}^{L-V+1=n}
v_{k+1}^{[k,2 \to k+1]}
{\tilde N}({\vec v}^{[k,2 \to k+1]},x)
= \sum_{{\vec v}^{\prime}}^{L-V+1=n} {v^{\prime}}_{k+1}
{\tilde N}({\vec v}^{\prime},x).
\end{equation}
Dropping the primes, we can thus write
\begin{eqnarray}
\label{deq}
& & \sum_{{\vec v}}^{L-V+1=n} \left[ \frac{v_2}{1 + x}
 + \sum_{k \geq 2} \frac{k (1 + (k - 1) x)}{1 + k x} v_{k+1}
\right] {\tilde N}({\vec v},x) \nonumber \\
& = & \sum_{{\vec v}}^{L-V+1=n} \left[ \sum_{k \geq 2}
v_k (k-1) - \sum_{k \geq 2} \frac{v_k (k-1) x}{1 + (k -1)x}
\right] {\tilde N}({\vec v},x) \nonumber \\
& = & \left( n - 1 - x \frac{\partial}{\partial x} \right)
\sum_{{\vec v}}^{L-V+1=n} {\tilde N}({\vec v},x) = 0,
\end{eqnarray}
which means
\begin{equation}
\sum_{{\vec v}}^{L-V+1=n} {\tilde N}({\vec v},x) =
C_n x^{n-1}, \ \ \ C_n = \sum_{{\vec v}}^{L-V+1=n} {\tilde N}({\vec v},1).
\end{equation}
Thus the sum formula has been proved up to a constant $C_n$.
\par
Let us then calculate $C_n$. First note that, according to
(\ref{nvx}), each ${\tilde N}({\vec v},x)$ contains only terms
of the order $x^V$ and lower orders. Due to the inequality
\begin{equation}
\label{ineq}
n - 1 - V = L - 2 V = \sum_{l=2}^{\infty} v_l (l - 2) \geq 0,
\end{equation}
this means that the largest order possible for a given
$n = L - V + 1$ is $x^{n-1}$. This order is reached only
for ${\vec v}$ with $v_3 = v_4 = \cdots = 0$, for which
the equality holds in (\ref{ineq}). Accordingly, we find
\begin{eqnarray}
\sum_{{\vec v}}^{L-V+1=n} {\tilde N}({\vec v},x) & = &
\sum_{{\vec v}}^{L-V+1=n} {\tilde N}({\vec v}) x^V
\prod_{l = 2}^{\infty} \left( l - 1 + \frac{1}{x} \right)^{v_l} \nonumber \\
& = & {\tilde N}((2)^{n-1}) x^{n-1} + {\rm lower \ order \ terms \ in} \ x.
\end{eqnarray}
Comparison with (\ref{deq}) now yields
\begin{equation}
C_n = {\tilde N}((2)^{n-1});
\end{equation}
all terms of lower orders in $x$ must mutually cancel.
In order to evaluate ${\tilde N}((2)^{n-1})$, we can utilize a
closed expression for ${\tilde N}_P({\vec v})$ (with 
$v_j \geq 0$ for $j \leq \Lambda$ and
$v_j = 0$ for $j > \Lambda$)
\begin{equation}
{\tilde N}_P({\vec v}) = \frac{(-1)^V}{L(L+1)}
\sum_{h_1 = 0}^{v_1} \sum_{h_2 = 0}^{v_2} \cdots
\sum_{h_{\Lambda} = 0}^{v_{\Lambda}}
(-1)^{\sum_{j=1}^{\Lambda} (j+1) h_j} \frac{\displaystyle
(\sum_{j=1}^{\Lambda} j h_j)!
(\sum_{j=1}^{\Lambda} j (v_j - h_j))!}
{\displaystyle  \prod_{j=1}^{\Lambda} [ h_j! (v_j - h_j)!]},
\end{equation}
which was derived by J{\"u}rgen M{\"u}ller\cite{JM}. Using the
identity
\begin{equation}
\int_0^{\infty} {\rm e}^{-s} s^j {\rm d}s = j!,
\end{equation}
we can rewrite J{\"u}rgen M{\"u}ller's formula as
\begin{equation}
\label{integral}
{\tilde N}_P({\vec v}) = \frac{(-1)^V}{L(L+1)} \int_0^{\infty} {\rm d}x
\int_0^{\infty} {\rm d}y \ {\rm e}^{-x} {\rm e}^{-y} \prod_{j=1}^{\infty}
\frac{(y^j - (-x)^j)^{v_j}}{v_j!},
\end{equation}
so that
\begin{eqnarray}
{\tilde N}((2)^{n-1}) & = & {\tilde N}_P((2)^{n-1}) \nonumber
\\ & = & \frac{(-1)^{n-1}}{2 (n-1)(2n-1)(n-1)!}
\int_0^{\infty} {\rm d}x \int_0^{\infty} {\rm d}y
\ {\rm e}^{-x} {\rm e}^{-y} (y^2 - x^2)^{n-1} \nonumber \\
& = &
\frac{(-1)^{n-1}}{4(n-1)(2n-1) (n-1)!}
\int_0^{\infty} {\rm d}s \int_{-s}^s {\rm d}t
\ {\rm e}^{-s} s^{n-1} t^{n-1} \nonumber \\
& = & \left\{ \begin{array}{ll} \displaystyle
\frac{(2 n - 3)!}{n!}, & n \ {\rm odd},
\\ 0, & n \ {\rm even} \end{array} \right.
\end{eqnarray}
($s = x + y$, $t = x - y$), which establishes
the desired result (\ref{sf2}).
\par
It is easy to check that J{\"u}rgen M{\"u}ller's formula holds
for ${\vec v}$'s with small $L-V$ (for example,
${\tilde N}_P((1)^1) = -1$).
Therefore, in order to prove it in general, it is sufficient to verify
that it fulfills the recursion relation (\ref{recur}).
For that purpose, we first define an ``average''
$\left\langle \cdots \right\rangle_{\vec v}$ of a function $f(x,y)$ as
\begin{equation}
\left\langle f(x,y) \right\rangle_{\vec v} =
\int_0^{\infty} {\rm d}x \int_0^{\infty}
{\rm d}y {\rm e}^{-x} {\rm e}^{-y} f(x,y)
\prod_{j=1}^{\infty} (y^j - (-x)^j)^{v_j}.
\end{equation}
Since ${\tilde N}_P({\vec v}) = 0$ if any of $v_j$ is negative,
(\ref{recur}) evidently holds if $v_l = 0$. Hence
we focus on the case $v_l \geq 1$. Then partial
integrations yield a relation
\begin{eqnarray}
& & \left\langle 1 \right\rangle_{\vec v} - l
\left\langle \frac{y^{l-1} + (-x)^{l-1}}{y^l - (-x)^l} \right\rangle_{\vec v}
\nonumber \\
& = & \left\langle 1 \right\rangle_{\vec v} - \left\langle
\frac{\partial}{\partial y} \left( \frac{y^l}{y^l - (-x)^l} \right)
- \frac{\partial}{\partial x} \left( \frac{(-x)^l}{y^l - (-x)^l} \right)
- l \frac{y^{2 l-1} - (-x)^{2 l-1}}{(y^l - (-x)^l)^2}
\right\rangle_{\vec v}
\nonumber \\
& = & \int_0^{\infty} {\rm d}x \int_0^{\infty} {\rm d}y
{\rm e}^{-x} {\rm e}^{-y} \nonumber \\ & \times &
\left[ \frac{y^l}{y^l - (-x)^l} \frac{\partial}{\partial y}
- \frac{(-x)^l}{y^l - (-x)^l} \frac{\partial}{\partial x}
- l \frac{y^{2 l-1} - (-x)^{2 l-1}}{(y^l - (-x)^l)^2}
\right] \prod_{j=1}^{\infty} (y^j - (-x)^j)^{v_j} \nonumber \\
\end{eqnarray}
for $l = 1,2,\cdots,L$. Using this relation and the identity
\begin{eqnarray}
& & \left[ \frac{y^l}{y^l - (-x)^l} \frac{\partial}{\partial y}
- \frac{(-x)^l}{y^l - (-x)^l}
\frac{\partial}{\partial x}
\right] \prod_{j=1}^{\infty}
(y^j - (-x)^j)^{v_j} \nonumber \\
& = & \sum_{k \geq 1} k v_k
\frac{y^{k+l-1} - (-x)^{k+l-1}}{(y^l - (-x)^l)(y^k - (-x)^k)}
\prod_{j=1}^{\infty} (y^j - (-x)^j)^{v_j},
\end{eqnarray}
we can readily derive
\begin{equation}
\label{rel1}
\left\langle 1 \right\rangle_{\vec v} - l
\left\langle \frac{y^{l-1} + (-x)^{l-1}}{y^l - (-x)^l} \right\rangle_{\vec v}
= \sum_{k \geq 1} k \left\langle (v_k - \delta_{kl})
\frac{y^{k+l-1} - (-x)^{k+l-1}}{(y^l - (-x)^l)(y^k - (-x)^k)}
\right\rangle_{\vec v}.
\end{equation}
\par
The following identity
\begin{eqnarray}
& & \int_0^{\infty} {\rm d}x \int_0^{\infty}
{\rm d}y \ {\rm e}^{- \omega x} {\rm e}^{- \omega y} \frac{1}{x + y}
\prod_{j=1}^{\infty} (y^j - (-x)^j)^{v_j} \nonumber \\
& = & \omega^{-L-1}
\int_0^{\infty} {\rm d}x \int_0^{\infty}
{\rm d}y \ {\rm e}^{-x} {\rm e}^{-y} \frac{1}{x + y}
\prod_{j=1}^{\infty} (y^j - (-x)^j)^{v_j}
\end{eqnarray}
can be proved by a transformation of the variables
$\omega x \mapsto x$,
$\omega y \mapsto y$. Differentiating the both sides of
this identity with respect to $\omega$ and then putting
$\omega = 1$, we obtain a relation
\begin{equation}
\frac{1}{L+1} \left\langle 1 \right\rangle_{\vec v} =
\left\langle \frac{1}{x + y} \right\rangle_{\vec v},
\end{equation}
from which it follows that
\begin{equation}
\frac{1}{L+1} \left\langle 1 \right\rangle_{\vec v}
= \left\langle
\frac{1}{y^l - (-x)^l} \left\{
y^{l-1} + (-x)^{l-1} - \frac{x y^{l-1} + y (-x)^{l-1}}{x + y}
\right\} \right\rangle_{\vec v}.
\end{equation}
Then, utilizing
\begin{equation}
\frac{xy^{l-1} + y (-x)^{l-1}}{x + y}
= - \frac{1}{2} \sum_{1 \leq m \leq l-2} \left[ (-x)^m y^{l - m - 1} + (-x)^{l
- m - 1}
y^m \right],
\end{equation}
we find
\begin{eqnarray}
\label{rel2}
& & - \frac{2}{L+1} \left\langle 1 \right\rangle_{\vec v} +
l \left\langle \frac{y^{l-1} + (-x)^{l-1}}{y^l - (-x)^l}
\right\rangle_{\vec v} \nonumber \\ & = &
\sum_{1 \leq m \leq l - 2}
\left\langle \frac{(y^m - (-x)^m)(y^{l-m-1}
- (-x)^{l-m-1})}{y^l - (-x)^l} \right\rangle_{\vec v}.
\end{eqnarray}
Adding the both sides of (\ref{rel1}) and (\ref{rel2}), we arrive at
\begin{eqnarray}
& & \frac{L-1}{L+1} \left\langle 1 \right\rangle_{\vec v}  =
\sum_{k \geq 1} k \left\langle (v_k - \delta_{kl})
\frac{y^{k+l-1} - (-x)^{k+l-1}}{(y^l - (-x)^l)(y^k - (-x)^k)}
\right\rangle_{\vec v} \nonumber \\
& + &  \sum_{1 \leq m \leq l - 2}
\left\langle \frac{(y^m - (-x)^m)(y^{l-m-1}
- (-x)^{l-m-1})}{y^l - (-x)^l} \right\rangle_{\vec v},
\end{eqnarray}
which gives the desired recursion relation (\ref{recur}) with J{\"u}rgen
M{\"u}ller's formula (\ref{integral}) substituted.

\section{The GOE to GUE Transition}\label{GOEtoGUE}
\setcounter{equation}{0}
\renewcommand{\theequation}{6.\arabic{equation}}

The equal-parameter correlation function $R(s;B,B)$ describes the
transition between the GOE and GUE universality classes as the
magnetic field $B$ increases from zero\cite{SN,BGOS,NS}.  In this
section, we shall reproduce Saito and Nagao's semiclassical
calculation\cite{SN} of the form factor (the Fourier transform of
$R(s;B,B)$) and further derive a sum formula analogous to (\ref{sf2})
as a conjecture.
\par
The RMT prediction of the form factor in this case is
derived from Pandey and Mehta's two-matrix model\cite{PM}. For
small $\tau$ ($0 \leq  \tau \leq 1$), it can be written as
\begin{eqnarray}
\label{pmfc}
K_{\rm RM}(\tau) & = & \tau + \frac{1}{2}
\int_{1-2\tau }^1 {\rm d}k \ \frac{k}{k + 2\tau}
{\rm e}^{-\mu(k + \tau)}
\nonumber \\ & = & \tau +{\rm e}^{-\mu}\tau+{\rm
e}^{-\mu}\left(\frac{\sinh \tau\mu}{\mu}-\tau\right) -
2 \tau^2 {\rm e}^{\mu(\tau-1)}\int_0^1\frac{{\rm e}^{-2\tau\mu
y}}{1+2\tau y} {\rm d}y. \nonumber \\
\end{eqnarray}
In the GOE limit the parameter $\mu$ is zero
and in the GUE limit it goes to infinity.
\par
The semiclassical argument is similar to that
in {\S} 3 and {\S} 4. The difference is that we have to take
account of the mutually time reversed pairs of loops and
segments of classical orbits. Following a similar
argument as in {\S} 3, we obtain a diagonal approximation
for the form factor
\begin{eqnarray}
K_{\rm PO}^{\rm (diag)}(\tau) &=& \tau + \tau {\rm e}^{- b T}\,,\\
b &=& 4 B^2 D/\hbar^2\,.
\end{eqnarray}
The RMT parameter $\mu$ should be equated with $b T$ in reference
to the semiclassical result.
\par
In order to extend the calculation to the off-diagonal terms, we need
to introduce integers $n_{{\rm enc},r}$ and $M$ characterizing the
structure of the orbit pairs as follows.  Let us fix an arbitrary
direction $(+)$ in which the orbits pass through the $r$-th encounter
and call the opposite direction $(-)$. Suppose that the orbit $\gamma$
passes through the encounter $\#^{(+)}(\gamma)$ and $\#^{(-)}(\gamma)$ times
in $(+)$ and $(-)$ directions, respectively. We then define the number
$n_{{\rm enc},r}$ as
\begin{equation}
n_{{\rm enc},r} = \frac{1}{2}
\left| \left\{ \#^{(+)}(\gamma)-\#^{(-)}(\gamma) \right\} -
 \left\{ \#^{(+)}(\gamma^{\prime}) - 
\#^{(-)}(\gamma^{\prime}) \right\} \right|.
\end{equation}
Moreover we define $M$ as the number of the pairs of mutually time
reversed loops.
\par
As before, for a general orbit pair $\alpha$ with $L$ loops
and $V$ encounters, the number of encounters in one periodic
orbit of a period $T$ is evaluated as
\begin{equation}
\int d{\bf u}d{\bf s}
\int_0^{T-t_{\alpha}} dT_1
\int_0^{T-t_{\alpha}-T_1} dT_2 \cdots
\int_0^{T-t_{\alpha}-T_1-T_2-\cdots-T_{L-2}} dT_{L-1}
\ Q_{\alpha},
\end{equation}
where
\begin{equation}
Q_{\alpha} = N(\mbox{\boldmath $v$},M) \frac{T}{L
\prod_{r=1}^V t_{{\rm enc},r} \ \Omega^{L-V}}.
\end{equation}
Here the combinatorial factor $N(\mbox{\boldmath $v$},M)$ depends on
a matrix $\mbox{\boldmath $v$}$ and $M$. The component $v_{lm}$
of the matrix $\mbox{\boldmath $v$}$ is the number of the encounters
with $l_r = l$ and $n_{{\rm enc},r} = m$. One can write
\begin{equation}
\mbox{\boldmath $v$} = (2,0)^{v_{20}} (2,1)^{v_{21}} (2,2)^{v_{22}} \cdots.
\end{equation}
Following the argument in \cite{MHBHA04,MHBHA05,SM}, we can identify
$N(\mbox{\boldmath $v$},M)$
with the number of generalized permutations satisfying
suitable conditions.
\par
Let us consider the effect of the gauge potential.
The Gaussian average (\ref{gaussian}) on the loops
gives a factor ${\rm e}^{-b T_1} {\rm e}^{-b T_2} \cdots
{\rm e}^{-b T_M}$, while from an encounter it yields
${\rm e}^{- b (n_{{\rm enc},r})^2 t_{{\rm enc},r}}$.
Thus we conclude that the total contribution to
the form factor from the orbit pair $\alpha$ is
\begin{eqnarray}
& & K_{{\rm PO},\alpha}(\tau)
= \tau \int d{\bf u}d{\bf s}
\nonumber \\ & \times  &
\int_0^{T-t_{\alpha}} dT_1
\int_0^{T-t_{\alpha}-T_1} dT_2 \cdots
\int_0^{T-t_{\alpha}-T_1-T_2-\cdots-T_{L-2}} dT_{L-1}
Q_{\alpha} R_{\alpha} e^{i \Delta S/\hbar} \nonumber \\
\end{eqnarray}
with
\begin{equation}
R_{\alpha} = e^{-b(T_1 + T_2 + \cdots + T_M)}
e^{-b((n_{{\rm enc},1})^2 t_{{\rm enc},1}
 + (n_{{\rm enc},2})^2 t_{{\rm enc},2} + \cdots +
(n_{{\rm enc},V})^2 t_{{\rm enc},V})}.
\end{equation}
This contributes to the terms of order $\tau^n$ with $n = L - V +
1$. As before we expand $K_{\rm PO,\alpha}(\tau)$ in $t_{{\rm
enc},r}$ and extract the term where all $t_{{\rm enc},r}$'s
mutually cancel. Then we find that the off diagonal contribution
to the form factor is
\begin{equation}
K_{\rm PO}^{\rm (off)}(\tau)
 = \sum_{\mbox{\boldmath $v$}} \sum_{M=0}^{L} N(\mbox{\boldmath $v$},M)
\frac{\tau^2}{L} \left( \frac{1}{T_H} \right)^{L-V-1}
\prod_{l=2}^{\infty} \prod_{m=0}^{\infty}
\left( - l \frac{\partial}{\partial T} -
m^2 b \right)^{v_{l m}} f(T,M),
\end{equation}
where
\begin{equation}
f(T,M) =  \int_0^T {\rm d}T_1
\int_0^{T-T_1} {\rm d}T_2 \cdots
\int_0^{T-T_1-T_2-\cdots-T_{L-2}} {\rm d}T_{L-1}
{\rm e}^{-b(T_1 + T_2 + \cdots T_M)}.
\end{equation}
If $M = 0$, the direction of motion along all loops and hence in
all encounters does not change in the partner orbit; consequently
$n_{{\rm enc},r}=0$ for all encounters. The corresponding
structures also exist in the case without time-reversal
invariance, so that 
\begin{equation}
N(\mbox{\boldmath $v$},0) = \left\{ \begin{array}{ll} 
N((2)^{v_{20}} (3)^{v_{30}} \cdots), & {\rm if \ all} \ v_{nj} 
\ {\rm with} \ j \neq 0 \ {\rm vanish}, \\ 
0, &  {\rm otherwise}. \end{array} \right.
\end{equation} 
Here $N(\vec v)$ is the number of structures introduced in 
{\S} 4 and {\S} 5. Time reversal of each such partner orbit 
produces another partner with $M=L$; therefore 
\begin{equation}
N(\mbox{\boldmath $v$},L) = \left\{ \begin{array}{ll} 
N((2)^{v_{22}} (3)^{v_{33}} \cdots), & {\rm if \ all} \ v_{nj} 
\ {\rm with} \ j \neq n \ {\rm vanish}, \\ 
0, &  {\rm otherwise}. \end{array} \right.
\end{equation} 
Note that the structures with $M = 0,L$ may exist only for odd $n=L-V+1$; 
see \cite{MHBHA05}.
\par
Noting the above relations for the combinatorial factors,
we can evaluate the contribution of the structures with $M=0,L$
in the same way as in {\S} 4, namely by Laplace-transforming the
corresponding summands in $K_{\rm PO}^{\rm (off)}(\tau)/\tau^2$,
using the sum rule (\ref{sf2}) for $N(\vec v)$ and transforming
back to the time representation. The contribution of the
structures with $M=0$ turns out to be zero whereas the structures with
$M=L$ reproduce the third summand in the last line of (\ref{pmfc}).
On the other hand, making the Laplace transform of the part of
$K_{\rm PO}^{\rm (off)}(\tau)/\tau^2$ with $1 \leq M \leq L-1$
and equating the result to the corresponding RMT prediction
deduced from the integral in the last line of (\ref{pmfc}),
we arrive at a conjecture
\begin{eqnarray}
\label{conj} & & \sum_{\mbox{\boldmath $v$}}^{L-V+1=n}
\sum_{M=1}^{L-1} \frac{N(\mbox{\boldmath $v$},M)}{L}
\frac{(-1)^V}{(1 + x)^M} \prod_{l=2}^{\infty} \prod_{m =
0}^{\infty}
(l + m^2 x)^{v_{l m}} \nonumber \\
& = & \frac{1}{(1 + x)^{n-1}} \sum_{p=1}^{n-1} \left(
\frac{x}{1 + x} \right)^{n - p - 1} (2 n - p - 3)!
\sum_{j=p}^{n-1} \frac{(-1)^j 2^j}{j (n - j - 1)!
(j - p)!} .\nonumber\\
\end{eqnarray}
In the cases $n = 2$ and $3$, the conjecture (\ref{conj}) was
substantially proved in \cite{SN}; by machine-assisted counting it
was verified up to $n = 7$. For small values of $n$ up to $4$, the
relevant $N(\mbox{\boldmath $v$},M)$'s are tabulated in Table 2.
Moreover, putting $x = 0$, we obtain
\begin{equation}
\sum_{\mbox{\boldmath $v$}}^{L-V+1=n} \sum_{M=1}^{L-1}
\frac{N(\mbox{\boldmath $v$},M)}{L} (-1)^V \prod_{l=2}^{\infty}
l^{v_l} = (-2)^{n-1} \frac{(n-2)!}{n-1}
\end{equation}
($v_l = \sum_{m=0}^{\infty} v_{lm}$), which is relevant 
to the GOE form factor. This special case was proved in 
\cite{MHBHA04,MHBHA05,SM}. The full proof of (\ref{conj}) 
is an interesting open problem.

\section*{Acknowledgements}

This work was partially supported by the Ministry of Education, 
Culture, Sports, Science and Technology, Government of Japan 
(KAKENHI 16740224) and by the Sonderforschungsbereich SFB/TR12 
of the Deutsche Forschungsgemeinschaft. The authors are grateful to 
Dr. J{\"u}rgen M{\"u}ller for providing his original 
result\cite{JM} before publication.

\newpage

\begin{table}[ht]
\begin{center}
\begin{tabular}{|c|c|c|c|c|}
\hline
$n$ & ${\vec v}$ & L & V & $N({\vec v})$ \\
\hline \hline
$3$ & $(2)^2$ & 4 & 2 & 1 \\
\cline{2-5} & $(3)^1$ & 3 & 1 & 1 \\
\hline \hline
$5$ & $(2)^4$ & 8 & 4 & 21 \\
\cline{2-5} & $(2)^2 (3)^1$ & 7 & 3 & 49 \\
\cline{2-5} & $(2)^1 (4)^1$ & 6 & 2 & 24 \\
\cline{2-5} & $(3)^2$ & 6 & 2 & 12 \\
\cline{2-5} & $(5)^1$ & 5 & 1 & 8 \\
\hline
\end{tabular}
\end{center}
\caption[table1]{The number $N({\vec v})$ of the
orbit structures corresponding to the
vector ${\vec v} = (2)^{v_2} (3)^{v_3} (4)^{v_4}
\cdots$, $L = \sum_l l v_l$, $V = \sum_l v_l$ and
$n = L - V + 1$.}
\end{table}

\begin{table}[ht]
\begin{center}
\begin{tabular}{|c|c|c|c|c|c|}
\hline
$n$ & $\mbox{\boldmath $v$}$ & L & V & M & $N(\mbox{\boldmath $v$},M)$ \\
\hline \hline
$2$ & $(2,0)^1$ & 2 & 1 & 1 & 2 \\
\hline \hline
$3$ & $(2,0)^2$ & 4 & 2 & 2 & 4 \\
\cline{2-6} & $(2,0)^2$ & 4 & 2 & 0 & 1 \\
\cline{2-6} & $(2,1)^2$ & 4 & 2 & 2 & 4 \\
\cline{2-6} & $(2,2)^2$ & 4 & 2 & 4 & 1 \\
\cline{2-6} & $(3,0)^1$ & 3 & 1 & 2 & 3 \\
\cline{2-6} & $(3,0)^1$ & 3 & 1 & 0 & 1 \\
\cline{2-6} & $(3,1)^1$ & 3 & 1 & 1 & 3 \\
\cline{2-6} & $(3,3)^1$ & 3 & 1 & 3 & 1 \\
\hline \hline
$4$ & $(2,0)^3$ & 6 & 3 & 1 & 6 \\
\cline{2-6} & $(2,0)^3$ & 6 & 3 & 3 & 10 \\
\cline{2-6} & $(2,0)^2 (2,1)^1$ & 6 & 3 & 2 & 12 \\
\cline{2-6} & $(2,0)^1 (2,1)^2$ & 6 & 3 & 3 & 36 \\
\cline{2-6} & $(2,0)^1 (2,1)^1 (2,2)^1$ & 6 & 3 & 4 & 12 \\
\cline{2-6} & $(2,2)^2 (2,0)^1$ & 6 & 3 & 5 & 6 \\
\cline{2-6} & $(2,0)^1 (3,0)^1$ & 5 & 2 & 1 & 15 \\
\cline{2-6} & $(2,0)^1 (3,0)^1$ & 5 & 2 & 3 & 15 \\
\cline{2-6} & $(2,0)^1 (3,1)^1$ & 5 & 2 & 2 & 25 \\
\cline{2-6} & $(2,0)^1 (3,2)^1$ & 5 & 2 & 3 & 10 \\
\cline{2-6} & $(2,0)^1 (3,3)^1$ & 5 & 2 & 4 & 5 \\
\cline{2-6} & $(2,1)^1 (3,0)^1$ & 5 & 2 & 2 & 20 \\
\cline{2-6} & $(2,1)^1 (3,1)^1$ & 5 & 2 & 3 & 20 \\
\cline{2-6} & $(2,2)^1 (3,1)^1$ & 5 & 2 & 4 & 10 \\
\cline{2-6} & $(4,0)^1$ & 4 & 1 & 1 & 12 \\
\cline{2-6} & $(4,0)^1$ & 4 & 1 & 3 & 4 \\
\cline{2-6} & $(4,1)^1$ & 4 & 1 & 2 & 16 \\
\cline{2-6} & $(4,2)^1$ & 4 & 1 & 3 & 8 \\
\hline
\end{tabular}
\end{center}
\caption[table1]{The number $N(\mbox{\boldmath $v$},M)$ of the
orbit structures corresponding to the matrix $\mbox{\boldmath $v$}
= (2,0)^{v_{20}} (2,1)^{v_{21}} (2,2)^{v_{22}} \cdots$, $L =
\sum_l \sum_m l v_{l m}$, $V = \sum_l \sum_m v_{l m}$ ,
$n = L - V + 1$ and the number $M$ of the pairs of mutually time
reversed loops. By machine-assisted counting the table is
extended to higher values of $n$.}
\end{table}

\end{document}